\documentclass[%
 reprint,
superscriptaddress,
 amsmath,amssymb,
 aps,
]{revtex4-2}

\usepackage{graphicx}
\usepackage{dcolumn}
\usepackage{bm}
\usepackage{braket} 
\usepackage[colorlinks,citecolor=blue,urlcolor=blue,linkcolor=blue]{hyperref}

\usepackage{color}
\usepackage{xcolor}
\usepackage{rotating}
\usepackage{mathrsfs}
\usepackage{amsthm}
\newtheorem{lemma}{Lemma}

\begin{document}

\preprint{APS/123-QED}

\title{Imaginary gap-closed points and dynamics in a class of dissipative systems}

\author{Shicheng Ma}
\email{First and second author contribute equally to this work}
\affiliation{School of Physics, Nankai University, Tianjin 300071, China}
\author{Heng Lin}
\email{linh1721@outlook.com}
\affiliation{Department of Physics, Tsinghua University, Beijing 100084, China}
\author{Jinghui Pi}
\email{pijh14@gmail.com}
\affiliation{Department of Physics, Tsinghua University, Beijing 100084, China}
\date{\today}

\begin{abstract}

We investigate imaginary gap-closed (IGC) points and their associated dynamics in dissipative systems. In a general non-Hermitian model, we derive the equation governing the IGC points of the energy spectrum, establishing that these points are only determined by the Hermitian part of the Hamiltonian. Focusing on a class of one-dimensional dissipative chains, we explore quantum walks across different scenarios and various parameters, showing that IGC points induce a power-law decay scaling in bulk loss probability and trigger a boundary phenomenon referred to as "edge burst". This observation underscores the crucial role of IGC points under periodic boundary conditions (PBCs) in shaping quantum walk dynamics. Finally, we demonstrate that the damping matrices of these dissipative chains under PBCs possess Liouvillian gapless points, implying an algebraic convergence towards the steady state in long-time dynamics.

\end{abstract}

\maketitle


\section{\label{sec:level1}Introduction}
The field of non-Hermitian physics has witnessed remarkable development over the past few decades \cite{RevModPhys.87.61, RevModPhys.88.035002,el2018non,ashida2020non,RevModPhys.93.015005,doi:10.1146/annurev-conmatphys-040521-033133}. In classical systems, non-Hermiticity is introduced through controlled gain-loss and leads to novel phenomena  beyond Hermitian counterparts \cite{PhysRevLett.100.103904,PhysRevLett.102.220402,ruter2010observation,PhysRevA.82.031801,PhysRevLett.106.093902,PhysRevLett.106.213901,regensburger2012parity,feng2013experimental,peng2014parity,doi:10.1126/science.1258004,doi:10.1126/science.1258480,PhysRevLett.112.203901,hodaei2017enhanced,chen2017exceptional}. In quantum systems,  effective non-Hermitian Hamiltonians arise from conditional dynamics under continuous monitoring and post-selection of null measurements \cite{lindblad1976generators,PhysRevLett.68.580,*Molmer:93,PhysRevA.45.4879,carmichael1993open,RevModPhys.70.101,doi:10.1080/00018732.2014.933502} or via the Feshbach projection formalism \cite{gamow1928quantentheorie,PhysRev.96.448,*FESHBACH1958357,*FESHBACH1962287,Rotter_2009,moiseyev2011non}. Realizations of non-Hermitian systems span diverse open quantum platforms \cite{peng2016anti,li2019observation,ren2022chiral,xiao2017observation,PhysRevLett.119.190401,*PhysRevLett.123.230401,dora2019kibble,*PRXQuantum.2.020313,ozturk2021observation,gao2015observation,doi:10.1126/science.aaw8205,PhysRevLett.127.090501,naghiloo2019quantum}. Notably, non-Hermitian Hamiltonians can exhibit peculiar properties, for instance, the existence of exceptional points \cite{Heiss2001,Berry2004,PhysRevE.69.056216,heiss2012physics,jing2017high}, which induce novel universality classes of phase transitions in non-Hermitian quantum systems \cite{PhysRevLett.80.5243,*Bender_2007,PhysRevB.58.16051,PhysRevB.81.033103,PhysRevX.4.041001,PhysRevLett.113.250401,ashida2017parity,PhysRevLett.119.040601,PhysRevLett.121.203001,PhysRevA.99.052118,*10.21468/SciPostPhys.7.5.069,PhysRevLett.123.123601,PhysRevResearch.2.033069,10.21468/SciPostPhys.12.6.194,PhysRevLett.128.010402,PhysRevX.13.021007}. Another unique feature of non-Hermitian systems is
the non-Hermitian skin effect (NHSE), namely the anomalous localization of an extensive number of bulk-band eigenstates at the edges \cite{PhysRevLett.116.133903,PhysRevLett.121.086803,PhysRevLett.121.136802,PhysRevLett.121.026808}. The NHSE plays a central role in the non-Hermitian
topological phases and reshapes convectional bulk-boundary correspondence \cite{PhysRevLett.102.065703, PhysRevB.84.205128,PhysRevB.84.153101,Schomerus:13,longhi2015robust,PhysRevLett.118.040401,PhysRevLett.118.045701,PhysRevLett.120.146402,PhysRevX.8.031079,kawabata2019topological,PhysRevX.9.041015,PhysRevB.99.235112,PhysRevLett.126.216407,PhysRevLett.124.056802,PhysRevLett.123.066405,PhysRevLett.123.206404,PhysRevLett.126.216405,Xiong_2018,PhysRevB.97.121401,PhysRevX.8.041031,PhysRevB.99.201103,PhysRevLett.122.076801,PhysRevLett.123.016805,PhysRevLett.123.066404,PhysRevB.103.085428,PhysRevLett.125.126402}. The topology origin of the NHSE is intimately linked to the point gap of non-Hermitian Bloch Hamiltonians \cite{PhysRevLett.124.086801,PhysRevLett.128.223903}. 

On the other hand, the NHSE also induces novel dynamical phenomena in open quantum systems. These include damping behavior and diverging relaxation time in Liouvillian dynamics \cite{PhysRevLett.123.170401,PhysRevLett.127.070402,PhysRevResearch.2.043167,PhysRevLett.125.230604,PhysRevResearch.4.023160}, directional amplification of signals \cite{PhysRevB.103.L241408,Wen:22}, self-healing of skin modes \cite{PhysRevLett.128.157601}, and directional invisibility of scattered wave-packet \cite{PhysRevLett.131.036402,zhang2022universal}. Intriguingly, non-Hermitian dynamics exhibit boundary condition independence in the thermodynamic limit \cite{PhysRevB.104.125435}. However, in finite-size systems, energy spectra under open boundary conditions (OBCs) show dramatic differences from those under periodic boundary conditions (PBCs) due to the NHSE \cite{PhysRevLett.121.086803,PhysRevLett.125.126402,PhysRevLett.124.086801}. Thus, one may question  which non-Hermitian Hamiltonians determine the dynamical evolution of the systems. Furthermore, the boundary's role is also crucial to understanding non-Hermitian dynamics.

Motivated by the above questions, we first analyze a general non-Hermitian model whose non-Hermiticity comes from the onsite dissipation. Our investigation centers on imaginary gap-closed (IGC) points, which are eigenstates possessing real energies and dictate the long-time behavior of the system \cite{xue_non-Hermitian_2022}. Notably, these IGC points depend solely on the Hermitian component of the Hamiltonian, unaffected by dissipative terms. Furthermore, IGC states are populated on non-dissipative sites and can be viewed as dark modes. When focusing on a class of one-dimensional dissipative models, the PBC spectra exhibit IGC points, while their OBC counterparts lack such points as eigenstates feature the NHSE. By introducing the non-Hermitian quantum walk, we investigate the relation between IGC points and the scaling behavior of bulk loss probability in space. Consequently, these IGC points lead to a spatial power-law decay of the loss probability in the bulk of the system. Moreover, an associated boundary phenomenon, termed "edge burst," emerges. Specifically, when a quantum particle is initially prepared on a specific bulk site, there is a prominent peak in the loss probability at the edge. We discuss this phenomenon in different quantum walk scenarios and the impact of various parameters, showing the universality of the correlation between edge burst and IGC points. 
Finally, we analyze the Liouvillian dynamics and demonstrate that damping matrices of these dissipative chains under PBCs can have Liouvillian gapless points, implying an algebraic convergence towards the steady state in long-time dynamics.      

This paper is organized as follows. In Sec.~\ref{Sec:2}, we introduce a general non-Hermitian model and analyze its IGC points. In Sec.~\ref{Sec:3}, we investigate the quantum walk dynamics in a class of one-dimensional dissipative chains. In Sec.~\ref{Sec:4}, we discuss the impact of various parameters on quantum walks. In Section~\ref{Sec:5}, we demonstrate that IGC points of the PBC Liouvillian spectrum capture the long-time dynamics of the system. Finally, a summary is given in Sec.~\ref{Sec:6}.

\section{Non-Hermitian Model and universal imaginary gapless modes\label{Sec:2}}

\begin{figure}[t]
\includegraphics[width=1\columnwidth]{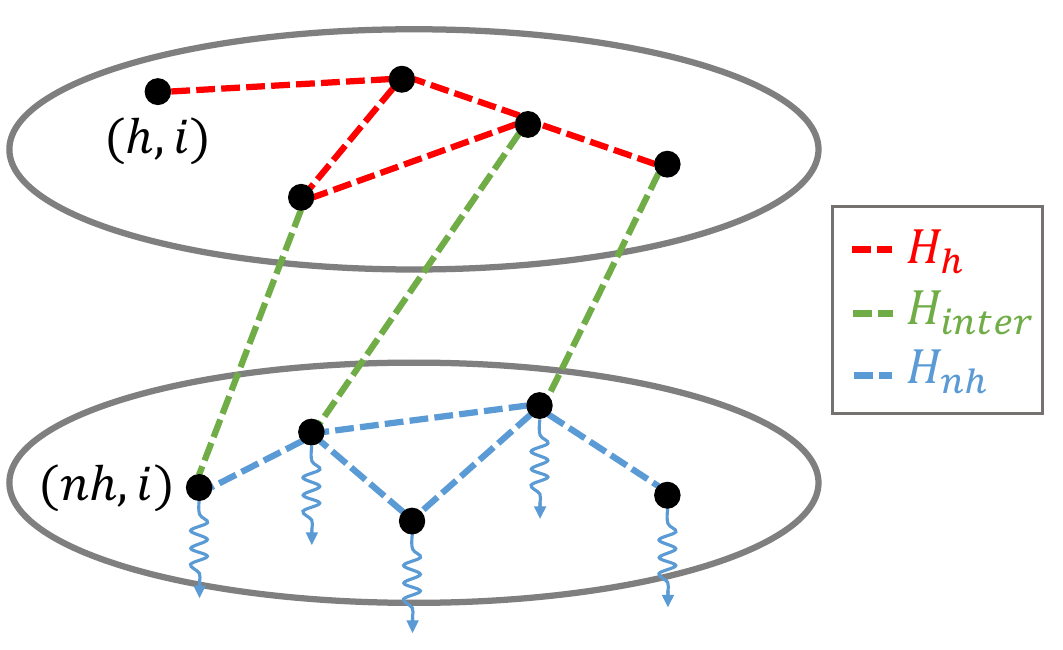}
	\caption{\label{FIG1}The general non-Hermitian lattice model. Red: interactions within non-dissipative sites; Blue: interactions within dissipative sites; Green: interactions between non-dissipative sites and dissipative sites.}
\end{figure}

Let us first consider a general non-Hermitian lattice model [Fig.~\ref{FIG1}], whose non-Hermiticity totally comes from the onsite dissipation. We classify the lattice sites into two categories: non-dissipative sites and dissipative sites, denoted by $(h, i)$ and $(nh, i)$, respectively. Then, the Hamiltonian of this system  can be divided into three parts,
\begin{equation}
	H = H_h + H_{nh} + H_{inter},
\end{equation}
where $H_h$ is the interactions within non-dissipative sites, $H_{nh}$ is the interactions within dissipative sites, and $H_{inter}$ is the interactions between non-dissipative sites and dissipative sites. Their expressions are given by
\begin{equation}
\begin{aligned}\label{eq:general_hamiltonian}
	H_h =& \sum_{i,j} a_{ij} \ket{h, i}\bra{h, j},\\
	H_{nh} =& \sum_{i,j} b_{ij} \ket{nh, i}\bra{nh, j},\\
	H_{inter} =& \sum_{i,j} \left(c_{ij}\ket{nh,i}\bra{h,j} + h.c.\right),
\end{aligned}
\end{equation}
where coefficients $a_{ij}$, $b_{ij}$ and $c_{ij}$ are complex number, satisfying $a_{ij}=a_{ji}^*$, $b_{ij}=b_{ji}^* (i\neq j)$ and $\mathrm{Im}(b_{ii})<0$ for all $i$ and $j$. Thus, $H_h$ and $H_{inter}$ are Hermitian operators, while $H_{nh}$ is non-Hermitian.

The imaginary part of the system's energy spectrum must be smaller than or equal to zero for its dissipative nature. Now we assume the existence of IGC points in the energy spectrum, i.e.
\begin{equation}\label{eq:conditions}
\begin{aligned}
		\exists \ket{IGC}, \quad s.t.\ \ &H\ket{IGC} = E_{IGC}\ket{IGC},\\ 
		&\text{and } Im(E_{IGC})=0.
\end{aligned}
\end{equation}
Subsequently, we discuss the conditions these IGC states should satisfy. Begin with the Lemma~\ref{lemma:1},
\begin{lemma}\label{lemma:1}
In systems where non-Hermiticity arises solely from onsite dissipation [Eq.~(\ref{eq:general_hamiltonian})], the wave functions of IGC states are only populated on non-dissipative sites, i.e.
\begin{equation}
	\ket{IGC} = \sum_{i} \psi_i^h \ket{h,i}.
\end{equation}
\end{lemma}

\begin{proof}
	Assume that
	\begin{equation}
		\ket{IGC} = \sum_{i} \psi_i^h \ket{h,i} + \sum_{j} \psi_j^{nh} \ket{nh,j}.
	\end{equation}
	Then 
	\begin{equation}\label{eq:proof1}
		\bra{IGC}(H - H^\dagger)\ket{IGC} = -2i\sum_j \gamma_j |\psi_j^{nh}|^2,
	\end{equation}
	where $-\gamma_j=\mathrm{Im}(b_{jj})$ is the dissipative rate of site $(nh, j)$, and $\gamma_j>0$. Notice that the LHS of Eq.~(\ref{eq:proof1}) also equals to
	\begin{equation}
		\bra{IGC}(H - H^\dagger)\ket{IGC} = E_{IGC}- E_{IGC}^* = 0.
	\end{equation}
	So $\psi_j^{nh} = 0$ for all $(nh, j)$ sites.
\end{proof}
Lemma~\ref{lemma:1} aligns well with our physical intuition. If an eigenstate has a population on dissipative sites, then the norm of the eigenstate will decrease with time, which implies the presence of an imaginary part in the energy. Via Lemma \ref{lemma:1}, the eigenequation of IGC states becomes
\begin{equation}\label{eq:eigenequation0}
\begin{aligned}
		0 =& (H-E_{IGC})\ket{IGC} \\
		  =& (H_h-E_{IGC})\ket{IGC} + H_{inter}\ket{IGC}.
\end{aligned}
\end{equation}
Notice that $(H_h-E_{IGC})\ket{IGC}$ is only populated on non-dissipative sites, while $H_{inter}\ket{IGC}$ is only populated on dissipative sites. Hence, Eq.~(\ref{eq:eigenequation0}) can be decomposed into two separate parts
\begin{equation}
	(H_h-E_{IGC})\ket{IGC} = 0,\quad H_{inter}\ket{IGC} = 0.
 \label{eq:condition}
\end{equation}
This implies that IGC states not only belong to the kernel of $H_{inter}$ but also function as eigenstates of a system solely comprised of non-dissipative sites, retaining their original eigenvalues. The term $H_{inter}$ can be interpreted as a representation of the coupling between non-dissipative and dissipative sites. Eq.~(\ref{eq:condition}) signifies that upon connecting the non-dissipative subsystem to the dissipative subsystem, these IGC states remain eigenstates of the combined system with their eigenvalues preserved.

In summary, IGC states can be described as dark modes characterized by the absence of a population on dissipative sites. Notably,  IGC states are the eigenstates of $H_h$, satisfying the connection conditions $H_{inter}$, independent of $H_{nh}$. Our findings hold true for a broad class of systems, as evidenced by their applicability to arbitrary Hamiltonian that only have non-Hermitian terms on its diagonal like Eq.~(\ref{eq:general_hamiltonian}). However, limitations arise when considering effective Hamiltonians with off-diagonal non-Hermitian terms. These non-diagonal forms, often encountered in systems like atoms coupled by the electromagnetic field \cite{skipetrov_absence_2014,skipetrov_magnetic-field-driven_2015,maximo_spatial_2015}, whose dissipative dynamics can lead to phenomena like fluorescence \cite{lo_dynamical_2020}, superradiance \cite{bellando_cooperative_2014,celardo_superradiance_2009,akkermans_photon_2008}, and subradiance \cite{guerin_subradiance_2016}. Our current theory necessitates further development to encompass these phenomena effectively.

\section{Imaginary Gapless modes and dynamics of Quantum walk} \label{Sec:3}
\begin{figure}[t]
\includegraphics[width=0.9\columnwidth]{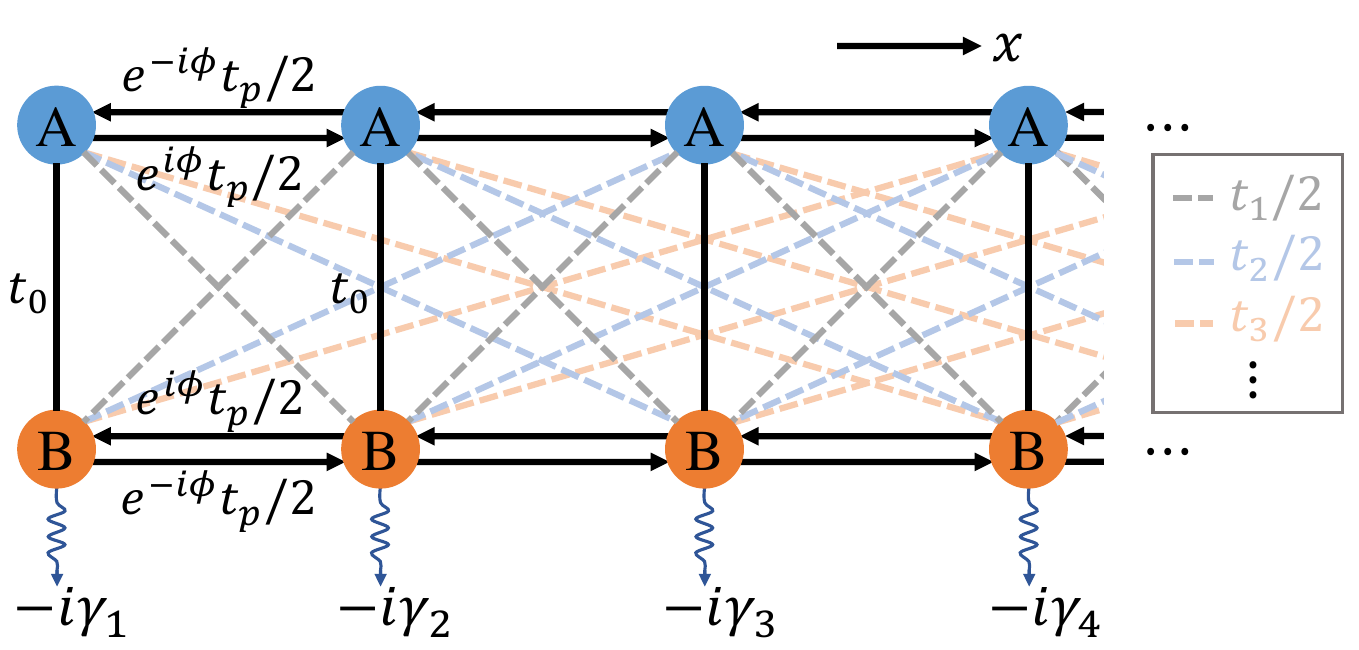}
\caption{\label{FIG2} Model of the one-dimensional dissipative ladder. Each unit cell is labeled by spatial coordinate $x$ and contains two sites, $A$ and $B$. The non-Hermiticity comes from site-dependent decay $\gamma_x$.}
\end{figure}

To investigate the impact of IGC modes on dynamical evolution, we delve into a more concrete model and analyze its quantum walk behavior. We consider a one-dimensional tight-binding model on a dissipative ladder of length $L$, as shown in Fig.~\ref{FIG2}. The A-B coupling range is $n$, meaning particles 
can hop between different sublattices up to the $n$-th nearest unit cells. Hopping along the same chain carries 
a Peierls phase $\phi$ \cite{goldman2016topological, PhysRevB.103.045402}, which induces gauge fluxes in closed hopping contours and breaks the time-reversal symmetry of the system. 
The non-Hermicity comes from the site-dependent dissipation $\gamma_x $, which induces the NHSE under OBC \cite{PhysRevLett.125.186802}. 
Then, we check the IGC states of this system under PBC. 
According to Eq.~(\ref{eq:condition}), IGC states of this system are also the eigenstates of chain A, given by
\begin{equation}\label{eq:eigeneqn_A}
		\ket{IGC} = \frac{1}{\sqrt{L}}\sum_{x=1}^L e^{ikx} \ket{x, A}.
\end{equation}
 These states possess eigenenergies
 \begin{equation}
 \quad E_{IGC} = t_p \cos{(k-\phi)},
 \label{IGCenergy}
 \end{equation}
where $k=2\pi n/L$ with $n=0,1,2,\cdots,L-1.$

The specific value of $k$ is determined by the connection condition obtained in Eq.~(\ref{eq:condition})
\begin{equation}\label{eq:connection_condition}
	\sum_{m=0}^{n}t_{m}\cos(mk)=0.
\end{equation}
At $k=0$, the LHS of Eq.~(\ref{eq:connection_condition}), denoted as $F(k)$, attains its maximum value $\sum_{m=0}^n t_m>0$. So, if the minimum value of $F(k)$ is smaller than or equal to zero, Eq.~(\ref{eq:connection_condition}) has real roots, and the energy spectrum is IGC. In finite chains, satisfying condition (\ref{eq:connection_condition}) with a discrete crystal momentum $k$ might not be achievable. However, from a dynamical perspective, their bulk dynamical behavior mirrors that of the infinite-chain limit due to the local nature of propagation. Treating $k$ as continuous in this limit enables the identification of rigorous IGC points.

\begin{figure}[b]
\includegraphics[width=1\columnwidth]{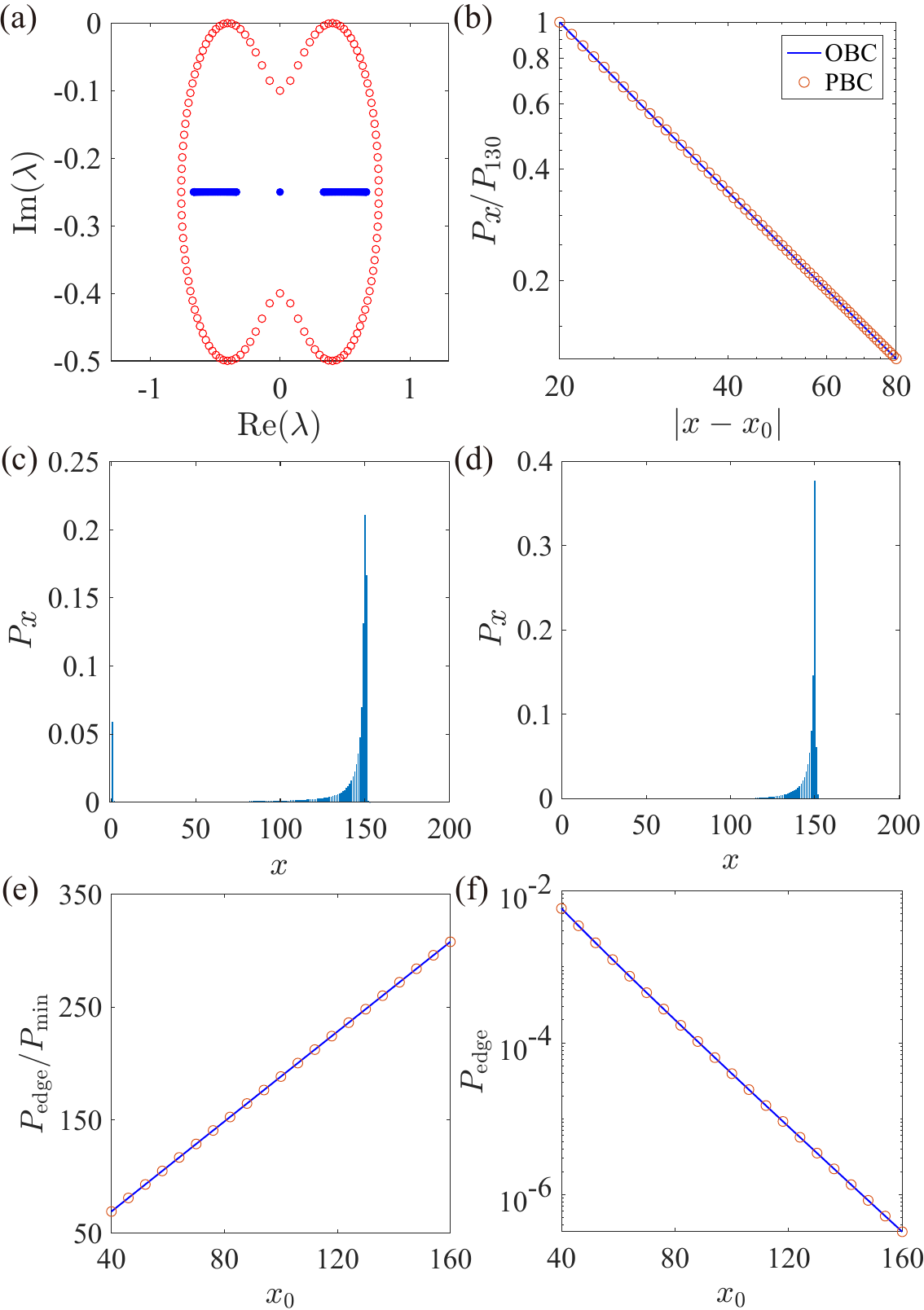}
\caption{\label{FIG3} The model with only the nearest A-B coupling (a) Energy spectra with $t_0=0.3$. Blue: OBC; Red: PBC. (b) The bulk distribution of the relative loss probabilities $P_x/P_{130}$ in a double logarithmic plot with $t_0=0.3$ for system size $L=200$, initial position $x_0=150$. (c), (d) The distribution of $P_x$ for a walker initiated at $x_0=150$ with $L=200$ under OBC. $t_0 = 0.3$ for (c) and $t_0 = 0.6$ for (d). (e) Relative height $P_{\rm{edge}}/P_{\min}$ and (f) the edge loss probability $P_{\rm{edge}}$ with chain length $L=200$, and $x_0$ varying from 40 to 160. $t_0 = 0.3$ for (c) and $t_0 = 0.6$ for (d) in a logarithmic plot. Throughout (a)-(f), $\phi=\pi/2$, $t_1=0.5$, $t_p=0.5$, and $\gamma=0.5$ are fixed. }
\end{figure}

Notably, the system under OBC lacks IGC states. This is because any IGC state would also be an eigenstate of the Hermitian $H_h$, which is an extended state, contradicting the NHSE's localized nature under OBC. Alternatively, in systems with the NHSE, OBC spectra are enclosed by their PBC counterparts \cite{PhysRevLett.124.086801, zhang2022universal}. For this dissipative lattice, the PBC spectra' maximum imaginary part is zero, rendering the OBC spectra always imaginary-gapped. An example is shown in Fig.~\ref{FIG3}(a).

Although substantial differences exist in the spectrum and eigenstates between OBC and PBC for this model, the dynamics within the bulk exhibit similarity for both cases. This similarity is evident from the Schrödinger equation $i(d/dt)\ket{\psi (t)}=H\ket{\psi (t)}$ (we set $\hbar=1$ throughout this paper). Under PBC, the Hamiltonian is given by $H_{PBC} = H_{OBC} + \delta H$. Here, $\delta H$ represents the coupling term that connects the two boundaries. When a wave function $\ket{\psi(t)}$ predominantly resides in the bulk, with negligible population at the boundaries, $\delta H \ket{\psi(t)} \approx 0$, and then consequently, the Schrödinger equations for OBC and PBC become effectively equivalent in this approximation. This feature is exemplified in Fig.~\ref{FIG3}(b). In the following, we show that the dynamical evolution properties of quantum walks under OBCs can be elucidated by IGC points of the PBC spectra. 

Considering a quantum walk in this model under OBC, a particle is released at the site $( x_0,A)$. Subsequently, the particle's wave function diffuses from the initial location $( x_0,A)$ and escapes from lossy $B$ sites. The wave function norm decreases as 
\begin{eqnarray}
      \frac{d}{dt}\left\langle \psi(t)|\psi(t)\right\rangle &&=i\left\langle \psi(t)|(H^{\dagger}-H)|\psi(t)\right\rangle\nonumber\\ 
      &&=-\sum_{x}2\gamma_x|\psi_{x}^{B}(t)|^{2}.
\end{eqnarray}
The escape probability from site $(x, B)$ is employed to characterize this quantum walk, given by
\begin{equation}
    P_x = 2\gamma_x \int_{0}^{\infty}{ |\psi_x^B(t)|^2 dt}.
\end{equation}
Then, we can use the non-Hermitian Green's function \cite{xue_non-Hermitian_2022} to express $P_x$ as
\begin{equation}
    P_x=\gamma_x\int_{-\infty}^{+\infty}\frac{d\omega}{\pi}\left| \braket{x,B\left| \frac{1}{\omega-H}\right|x_0, A} \right|^2.
    \label{eq:pxgf}
\end{equation}
Note that $\Sigma_{x}P_{x}=1$ is satisfied under the initial-state normalization $\left\langle \psi(0)|\psi(0)\right\rangle =1$.
Intuitively, the distribution $P_x$ features a peak centered on the site $x_0$, displaying left-right asymmetry attributable to the NHSE, which causes the wave function to move leftward preferentially \cite{liu_information_2021}. However, under certain parameters, a prominent peak, named edge burst, emerges at the left edge \cite{Wang_2021}. In a recent paper, Xue et al. investigated this model with the nearest A-B hopping term and uniform dissipation, i.e., $\gamma_x$ is a constant, revealing that the edge burst phenomenon results from the interplay between the IGC spectrum under PBC and the NHSE \cite{xue_non-Hermitian_2022}.

 When there are IGC states, the escape probability $P_x$ decays slowly as a power law with the distance from the initial position $x_0$,
\begin{equation}
    P_{x}\sim|x-x_{0}|^{-\alpha_{b}}.
    \label{eqpower}
\end{equation}
In this case, the walker remains a large wave function amplitude when it arrives at the left edge \cite{PhysRevA.109.022236}. Then, the walker becomes trapped due to the NHSE and escapes from  $(1,B)$ site over time, leading to a high peak $P_{\rm{edge}}$, as shown in Fig.~\ref{FIG3}(c). If we shift the left boundary to infinite, the walker is no longer localized at $x=1$ and carries out left walking quickly through this position, resulting in a tiny $P_{1}$. This argument gives an estimation of $P_{\rm{edge}}$ as follows,
\begin{equation}
    P_{\rm{edge}}\sim\sum_{-\infty}^{0}P_{x}\sim\int_{-\infty}^{0}\left\vert 
    x-x_{0}\right\vert ^{-\alpha_{b}}dx\sim\left(
    x_{0}\right)  ^{-\alpha_{b}+1}.
\label{eqpedge}
\end{equation}
Here, we have assumed that bulk $P_x$ is only determined by the relative distance to the initial position $x_0$, and system size effects can be neglected. On the other hand, Eq.~(\ref{eqpower}) implies that $P_x$ takes the minimum near the edge, which gives the estimation $P_{\min}\sim x_0^{-\alpha_{b}}$. Therefore, it follows from Eq.~(\ref{eqpedge}) that 
\begin{equation}
 P_{\rm{edge}}/{P_{\min}}\sim x_{0},
\end{equation}
which can be confirmed by our numerical results shown in Fig.~\ref{FIG3}(e). Thus, if the initial position $x_0$ is far from the edge, there is a large peak $P_{\rm{edge}}$ compared to almost invisible $P_x$  in the vicinity sites.

In contrast, when there are no IGC states, bulk $P_x$ decays fast as exponential law with 
\begin{equation}
 P_{x}\sim (\lambda_b)^{x_0-x},(\lambda_b<1).
\end{equation}
In this case, wave function amplitude becomes very small when the walker arrives at the left edge. Similarly, we have the estimation 
\begin{equation}
P_{\rm{edge}}\sim\sum_{-\infty}^{0}P_{x}\sim\int_{-\infty}^{0} (\lambda_b)^{x_0-x} dx\sim\left(
\lambda_b\right)  ^{x_0}.
\label{eqpedge2}
\end{equation}
$P_{\rm{edge}}$ is of the same order as the decay tail, and therefore, as shown in Fig.~\ref{FIG3}(d), no edge burst exists. The relationship between $P_{\rm{edge}}$ and $x_0$ in Eq.~(\ref{eqpedge2}) is verified by the numerical results shown in Fig.~\ref{FIG3}(f).

\section{Impact of parameters}\label{Sec:4}

The above discussion shows that IGC points of PBC spectra are crucial to the non-Hermitian dynamics.  In this section, we explore the quantum walk across various parameters, demonstrating the universality of the correlation between edge burst and IGC points. This correlation remains independent of the values or distribution of A-B hopping terms, Peierls phases, and loss rates. This indicates that the bulk dynamics can be effectively captured in these models by specific spectra characteristics under PBCs.

\subsection{A-B coupling model 2-nd nearest intercell hopping}
\begin{figure}[t]
\includegraphics[width=1\columnwidth]{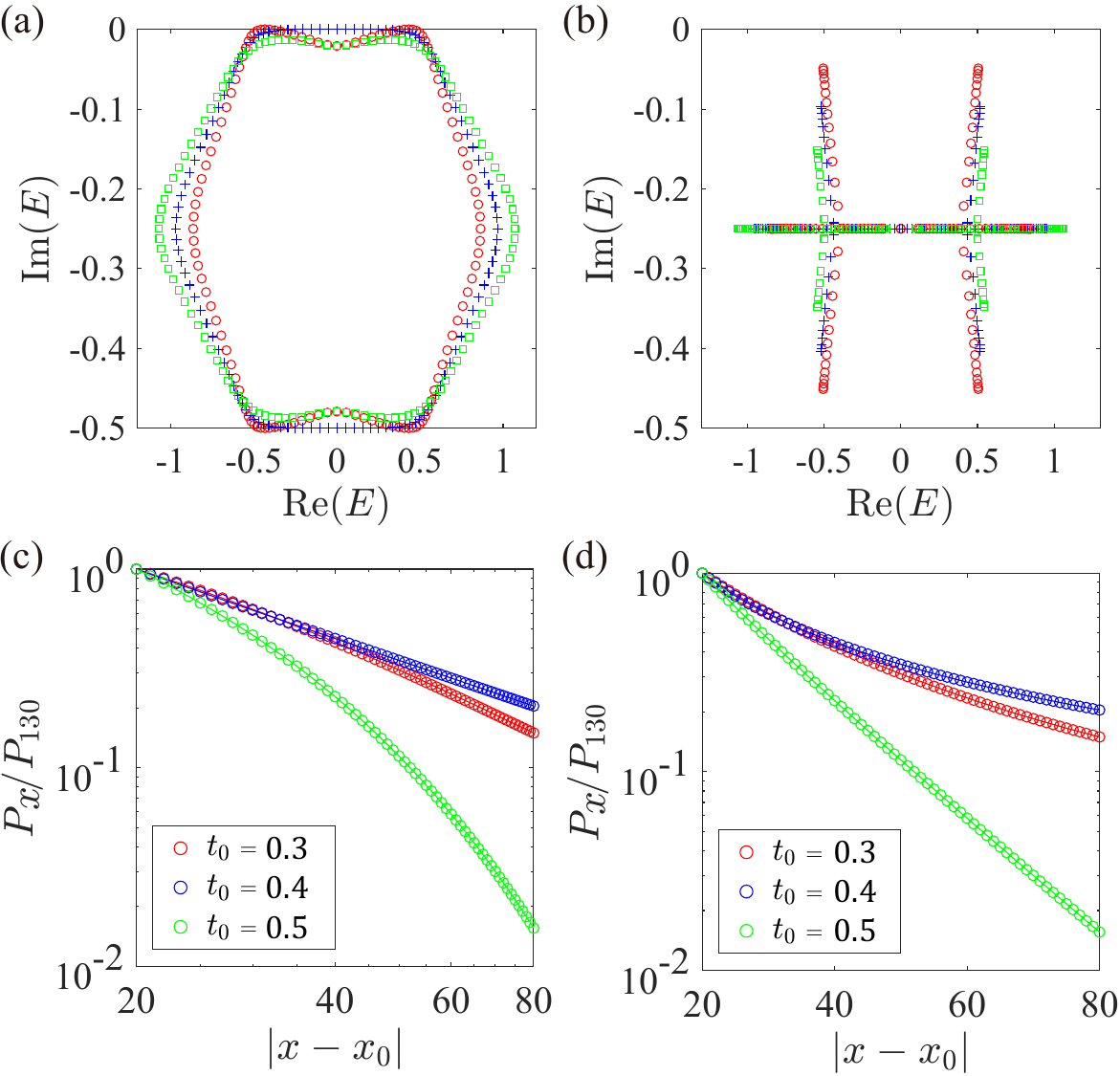}
\caption{\label{FIG4} (a), (b) Energy spectra 
  under (a) PBCs and (b) OBCs, $t_0=0.3$ (red), $0.4$ (blue), and $0.5$ (green). (c), (d) The bulk distribution of $P_x$ in double logarithmic (c) and logarithmic (d) plots. System size $L=200$, and the initial position $x_0=150$. Through (a)-(d), $t_p=0.5$, $t_1=0.5$, $t_2=0.1$, $\gamma=0.5$ are fixed. }
\end{figure}
\begin{figure}[t]
\includegraphics[width=1\columnwidth]{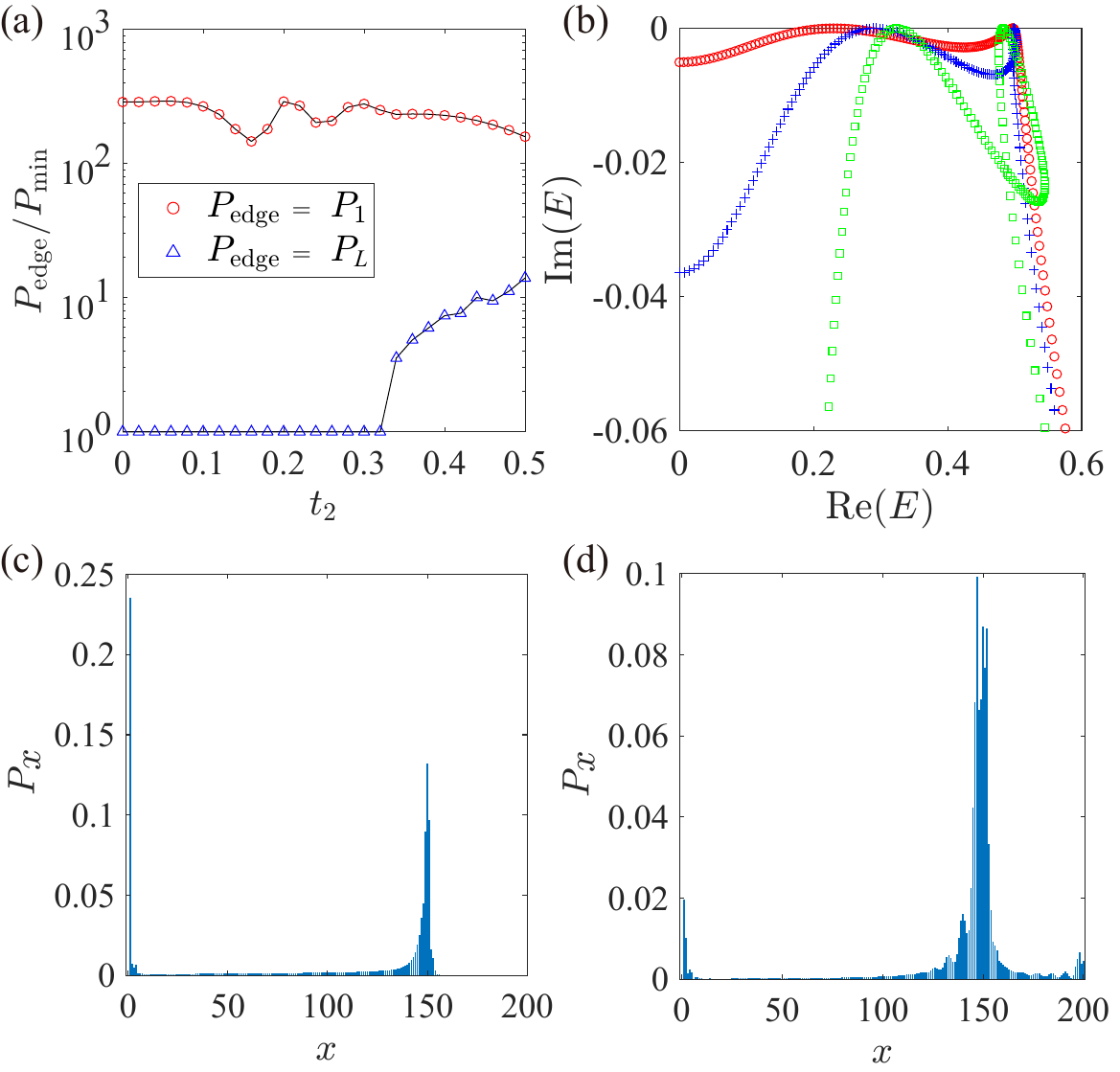}
\caption{\label{FIG5}  (a) The relative height $P_\mathrm{edge}/P_\mathrm{min}$ with varying $t_2$. (b) Part of the energy spectra with system size $L=500$ under PBCs, $t_2=0.25$ (red), $0.33$ (blue), and $0.50$ (green) (c), (d) The loss probability $P_x$ for a walker initiated at $x_0=150$ with $L=200$. $t_2=0.2$ for (c) and $t_2=0.5$ for (d). Through (a)-(d), $t_p=0.5$, $t_0=0.3$, $t_1=0.5$, $\gamma=0.5$ are fixed.}
\end{figure}
In this subsection, we consider the impact of A-B hopping terms with long range. For concreteness, the intercell A-B coupling extends to the second nearest neighbor, and the Peierls phase takes $\phi=\pi/2$. Additionally, a uniform loss rate of $\gamma$ is applied. The Bloch Hamiltonian is given by 
\begin{equation}
H(k) =h_x(k)\sigma_{x}+\left( h_y (k)+i\frac{\gamma}{2}\right)  \sigma_{z} 
\\-i\frac{\gamma}{2}I,
\label{eq20}
\end{equation}
where $h_x(k)= t_0 + t_1 \cos k + t_2 \cos 2k$, $ h_y (k) = t_{p}\sin k$, and $\sigma_{x,z}$ are the Pauli matrices.
Since $h_x(k)=F(k)$, the IGC points are determined by  $h_x(k)=0$. This condition, expressed in terms of the Bloch phase factor $\beta=e^{ik}$, becomes 
\begin{equation}
t_0 + \frac{t_1}{2} (\beta+\beta^{-1})+ \frac{t_2}{2} (\beta^{2}+\beta^{-2})=0.
\label{eq22}
\end{equation}
This is a quartic equation of $\beta$ with the constraint $\vert \beta \vert =1$. Therefore, the maximum number of IGC points is four, which is confirmed by the numerical results of the energy spectra. Furthermore, we can infer that if intercell A-B couplings are up to the $n$-th nearest neighbors, the maximum number of IGC points is $2n$. 

As $h_x(0)=t_0+t_1+t_2>0$ (maximum value of $h_x(k)$), the existence of IGC points solely depends on the minimum value, $h_x(k_0)$. By analyzing $h_x(k)$, we find that the expression for $h_x(k_0)$ takes a different form when the parameters change. 
When $ t_2 \leq t_1/4 $,  $h_x(k_0)$ is given by 
\begin{equation}
    h_x(k_0) =t_0-t_1 +t_2,
    \label{eq:m1}
\end{equation} 
with $k_0= \pi$. Conversely, for $t_2 > t_1/4 $, the expression changes to   
\begin{equation}
    h_x(k_0)=t_0-\frac{t_1^2}{8t_2}-t_2, \label{eq:m2}
\end{equation}
with $\cos k_0=-t_1/4t_2 $. Combining Eq.~(\ref{eq:m1}) and Eq.~(\ref{eq:m2}), we obtain the maximum value of $h_x(k_0)$ as $t_0-(t_1/\sqrt{2})$, achieved when $t_2=t_1/2\sqrt{2}$. Thus, the inequality $h_x(k_0)\leq 0$ always holds in the regime $ t_0 \leq t_1/\sqrt{2}$, regardless of the value of $t_2$. Consequently, the energy spectra of the Bloch Hamiltonian Eq.~(\ref{eq20}) are guaranteed to exhibit IGC points. However, for $ t_0 > t_1/\sqrt{2}$, the existence of IGC points becomes contingent on specific values of hopping parameters $t_0$, $t_1$, and $t_2$. For example, as shown in Fig.~\ref{FIG4}(a), when fixing other parameters and increasing the parameter $t_0$, two IGC points gradually close together. At a critical value, they merge into a single IGC point. Besides, no IGC points exist when $t_0$ exceeds this critical value. Apparently, the corresponding OBC spectra do not have IGC points, which is shown in Fig.~\ref{FIG4}(b). As discussed in Sec.~\ref{Sec:3}, whether energy spectra have IGC points can be linked to the quantum walk behavior. Fig.~\ref{FIG4}(c) and \ref{FIG4}(d) illustrate this connection, where the bulk distribution of particle loss probabilities $P_x$
exhibits power-law behavior for PBC spectra with IGC points and exponential behavior for spectra without them.

To demonstrate the feature of the edge burst phenomenon for this model, we choose specific parameter values $t_0=0.3$ and $t_1=0.5$. As previously established, the IGC condition is automatically satisfied because the maximum value of $h_x(k_0)$ is strictly negative. The relative heights $P_\mathrm{edge}/P_\mathrm{min}$ are vital to depict edge burst, as these physical quantities are closely related to the decay scaling behavior of the walker in the bulk. Here, $P_\mathrm{edge}$ are the escape probabilities at two edges and equal to $P_1$ or $P_L$.  $P_\mathrm{min}$ is defined as $\min\{P_1,P_2,\dots,P_{x_0}\}$ for $P_{\rm{edge}}=P_1$ and $\min\{P_{x_0},P_{x_0+1},\dots,P_{L}\}$ for $P_{\rm{edge}}=P_L$. We calculate $P_\mathrm{edge}/P_\mathrm{min}$ as the parameter $t_2$ increasing from zero [see Fig.~\ref{FIG5}(a)]. For small $t_2$, the ratio $P_1/P_{\min}>>1 $ while $P_L/P_{\min}\sim1$, which indicates a single remarkable probability loss peak at the left edge [Fig.~\ref{FIG5}(c)]. As $t_2$ exceeds a critical value,  $P_L/P_{\min}$ increases rapidly and becomes much larger than one. In this case, the system exhibits two remarkable probability loss peaks at both left and right edges [Fig.~\ref{FIG5}(d)]. This phenomenon is closely linked to the bipolar skin effect, which corresponds to the self-intersecting point of the PBC spectrum \cite{PhysRevB.106.235411} [see green line in Fig.~\ref{FIG5}(b)]. Another interesting aspect of the edge burst phenomenon is that the ratio of $P_1/P_{x_0}$ can be greater than 1 [see Fig.~\ref{FIG5}(c)], which is not observed in the system for $t_2=0$.

\subsection{Model with varying Peierls phases $\phi$}

\begin{figure}[b]
\includegraphics[width=1\columnwidth]{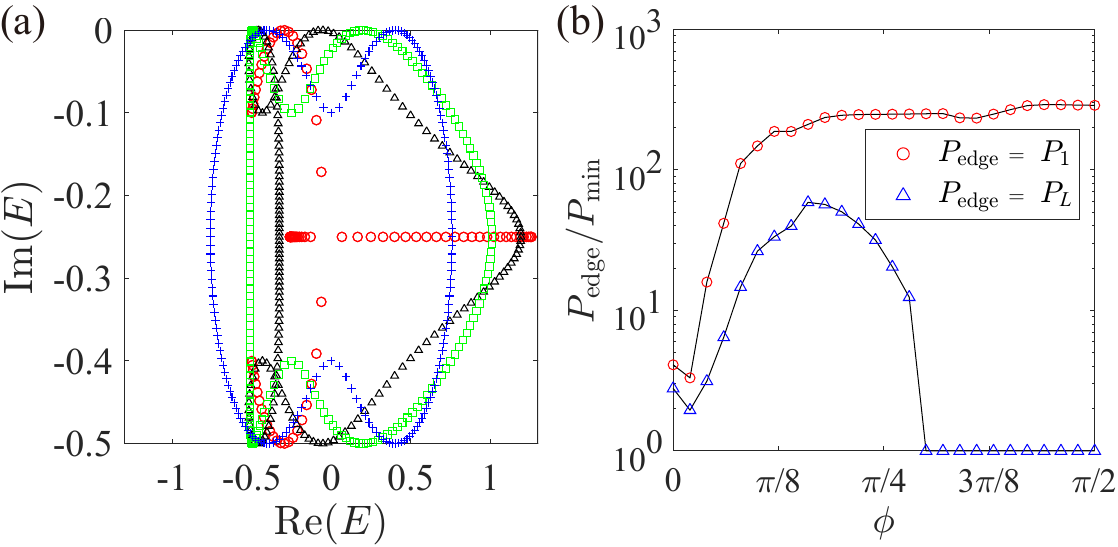}
\caption{\label{FIG6} (a) Energy spectra under PBCs. 
 Peierls phases $\phi$ are set to $0$ (red), $\pi/6$ (black), $\pi/3$ (green), and $\pi/2$ (blue). (b) The relative height $P_\mathrm{edge}/P_\mathrm{min}$ with varying $\phi$. System size $L=200$ and initial position $x_0=150$. Throughout (a)-(b), $t_p=0.5$, $t_0=0.3$, $t_1=0.5$, $\gamma=0.5$ are fixed.}
\end{figure}

In this subsection, we focus on the influence of the Peierls phase $\phi$. For simplicity, we only consider the nearest neighbor intercell A-B coupling with uniform loss rate $\gamma$, and the Bloch Hamiltonian is
\begin{equation}
    H(k)=(t_{0}+t_{1}\cos k)\sigma_{x}+\left[t_{p}\cos(k-\phi)+i\frac{\gamma}{2}\right]\sigma_{z}-i\frac{\gamma}{2}I.
\label{eq26}
\end{equation}
IGC points are given by $t_{0}+t_{1}\cos k=0$. The eigen
values of IGC states described by Eq.~(\ref{IGCenergy}) can be expressed in terms of the system parameters as follows
\begin{equation}
    E_{IGC}=\frac{t_p}{t_1}\left(-t_{0}\cos\phi\pm\sqrt{t_{1}^{2}-t_{0}^{2}}\sin\phi\right).
\end{equation}
This analytical expression exhibits good agreement with numerical results of energy spectra under PBCs, as shown in Fig.~\ref{FIG6}(a). Therefore, Peierls phases  $\phi$ do not alter the IGC conditions or impact the bulk scaling behavior of $P_x$. they only shift the values of $E_{IGC}$ along the real axis.

When $\phi=0$, the Hermitian part of the Bloch Hamiltonian Eq.~(\ref{eq26}) has time-reversal symmetry. Consequently, the onsite dissipation does not induce the NHSE. In this case, the distribution of $P_x$ is left-right symmetric, and there is no edge burst phenomenon. As $\phi$ increases from zero, this model breaks time-reversal symmetry and thus features the NHSE. We can see from Fig.~\ref{FIG6}(b) that the quantum walk exhibits a bipolar edge burst for a not very large $\phi$ and reduces to a standard edge burst on the left side when it exceeds some critical value.
 The corresponding PBC spectra also exhibit the emergence and disappearance of self-intersecting points, as shown in Fig.~\ref{FIG6}(a).

\subsection{Model with varying loss rate $\gamma_x$}

\begin{figure}[t]
\includegraphics[width=1\columnwidth]{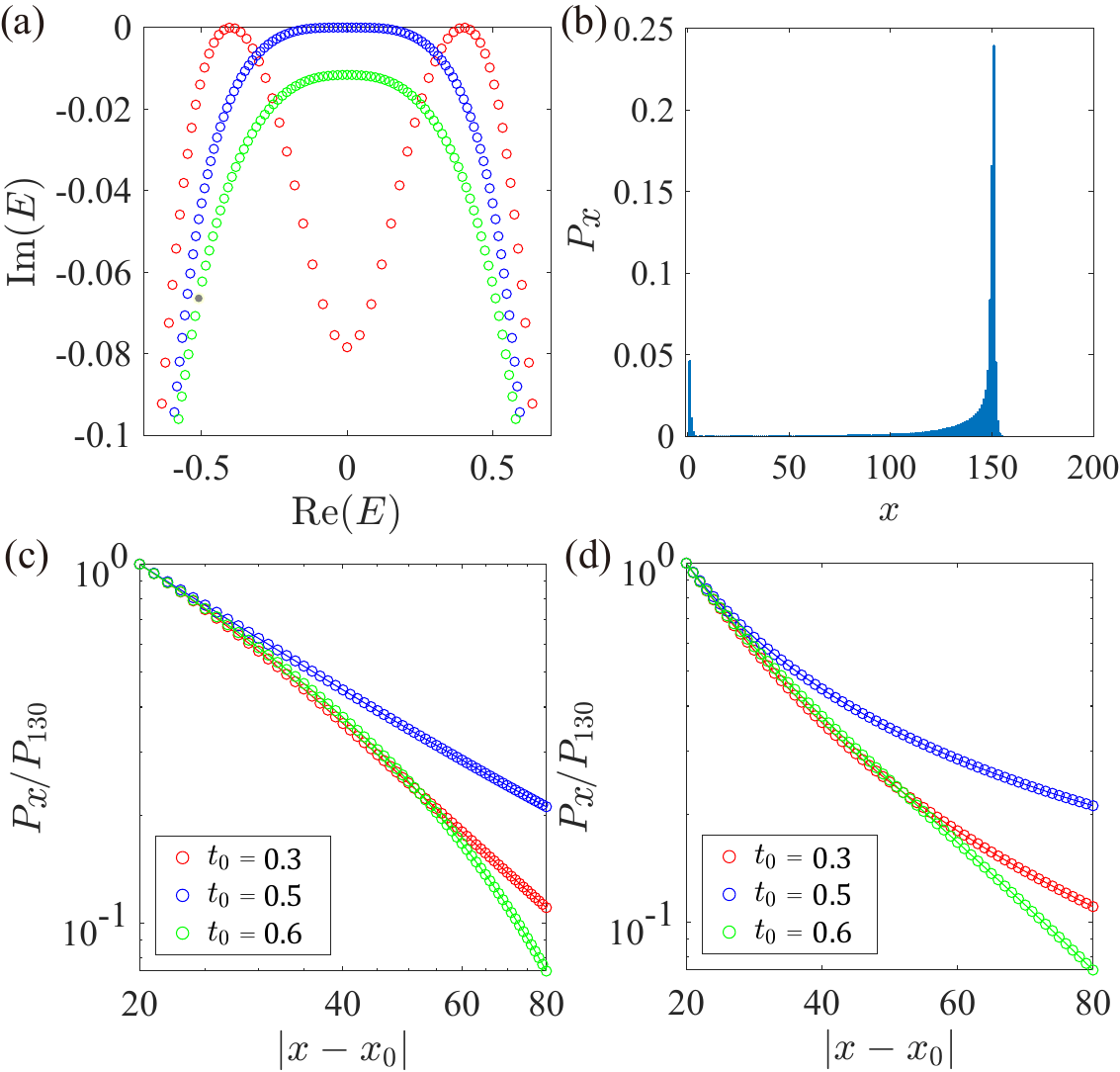}
\caption{\label{FIG7} (a) Energy spectra under PBCs. $t_0=0.3$ (red), $t_0=0.5$ (blue), $t_0=0.6$ (green).  (b) The distribution of $P_x$ when $t_0=0.3$. (c), (d) The bulk distribution of $P_x$ in double logarithmic (c) and logarithmic (d) plots with non-uniform loss rates. For (b)-(d), system size $L=200$ and the initial position $x_0 = 150$. Throughout (a)-(d), $t_p=0.5$, $t_1=0.5$, $\gamma_0=0.01$, $\gamma_c=0.20$ are fixed.}
\end{figure}

In this subsection, we discuss the effect of loss rate distribution. We focus on the model with nearest-neighbor intercell A-B coupling and $\phi=\pi/2$ for convenience. The system breaks discrete translational symmetry if the loss rate $\gamma_x$ is not uniform. Thus, the Hamiltonian under PBC cannot be expressed in a simple Bloch form, and eigenstates cease to be Bloch waves. However, as established in Sec.~\ref{Sec:2}, IGC states exhibit invariance, remaining unaffected by the specific form of $\gamma_x$. Let us consider a linear form of $\gamma_x$ as follows   
  \begin{equation}
    \gamma_x=\gamma_0x+\gamma_c.   \label{eq:23}
 \end{equation}
 As shown in Fig.~\ref{FIG7}(a), this non-uniform system transitions from IGC to imaginary gapped when increasing the value of $t_0$. IGC points of the PBC spectra also determine the bulk dynamics of quantum walk in this non-uniform system. When IGC points exist, $P_x$ follows a power-law decay in the bulk. Conversely, in the absence of points, $P_x$ follows an exponential law decay in the bulk. We can see this clearly in Fig.~\ref{FIG7}(c) and Fig.~\ref{FIG7}(d). Besides, a remarkable probability loss peak appeared at the edge when bulk $P_x$ shows a power-law decay, as shown in Fig.~\ref{FIG7}(b). This gives an explicit explanation of the edge burst phenomenon in Ref.~\cite{yuce_non-Hermitian_2023}.

\section{Relation to the Liouvillian dynamics }\label{Sec:5}
\begin{figure}[b]
\includegraphics[width=1\columnwidth]{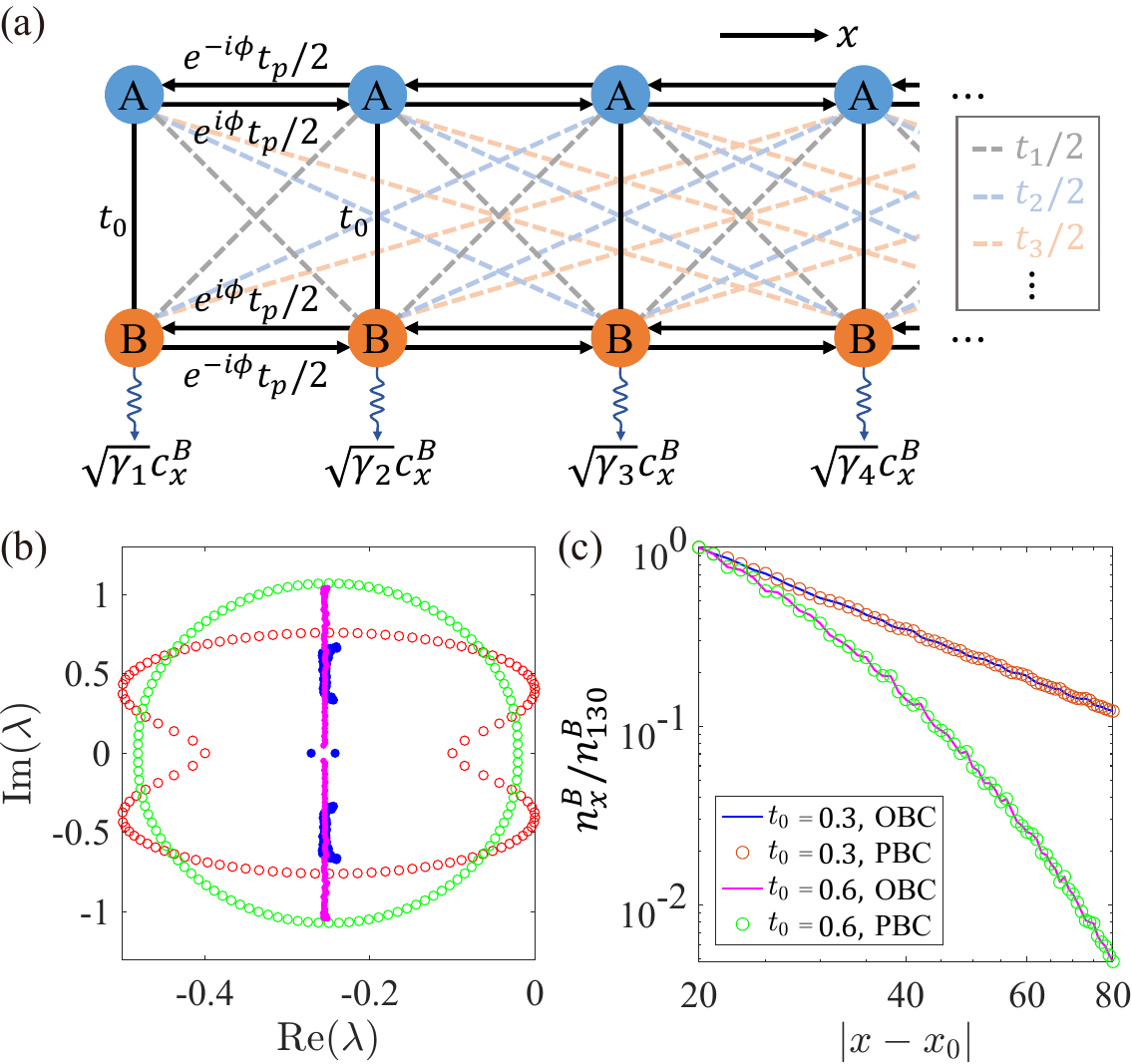}
\caption{\label{FIG8} (a) The corresponding bosonic quadratic open quantum system. (b), (c) The model only considers the nearest A-B coupling and chooses $\phi=\pi/2$, $t_1=0.5$, $t_p=0.5$, and $0.4<\gamma_x<0.6$ (randomly distributed). (b) Liouvillian spectra. Blue: OBC, $t_0=0.3$; Red: PBC, $t_0=0.3$; Purple: OBC, $t_0=0.6$; Green: PBC, $t_0=0.6$. (c) The bulk distribution of $P_x$ in a double logarithmic plot.}
\end{figure}
Leveraging the established correspondence between non-Hermitian dynamics and steady states of open quantum systems (as explored in Ref.~\cite{PhysRevB.108.235422}), our conclusions can be generalized to open quantum systems as well. When a quantum system is coupled to a Markovian bath, the corresponding dynamical evolution of the density matrix is governed by the master equation, which can be expressed in Lindblad formalism \cite{lindblad1976generators}
\begin{equation}
\frac{d\rho}{dt}   =-i\left[  H_0,\rho\right]  +\sum_{\alpha}\left(  2L_{\alpha}\rho
L_{\alpha}^{\dagger}-\left\{  L_{\alpha}^{\dagger}L_{\alpha},\rho\right\}  \right), 
\end{equation}
where $H_0$ is the Hamiltonian of the system, $L_{\alpha}$ are the Lindblad dissipators describing quantum jumps randomly moving the state $ \left\vert\psi \right\rangle$ to $ L_{\alpha} \left\vert \psi \right\rangle$. In this system, the quantum dynamics started from single particle states are controlled by the effective Hamiltonian $\mathcal{H}_{eff}=H_0 -i\sum_{\alpha}L_{\alpha}^{\dagger}L_{\alpha}$ as $d\rho / dt=i(\rho \mathcal{H}_{eff}^{\dagger }- \mathcal{H}_{eff} \rho)$.

In the open chain, we consider [Fig.~\ref{FIG8}(a)], the dissipators are given by $L_x=\sqrt{\gamma_x} c_{x}^\mathtt{B}$. Notably, the long time evolution of the system can be captured by the single-particle correlation function $C_{xy}(t)=\mathtt{Tr}\left[  \rho\left(  t\right)  c_{x}^{\dagger}c_{y}\right]$. For this dissipative system, any initial state will converge to the non-equilibrium steady state correlation $C(\infty)$, determined by  $dC(\infty)/dt =0$. In this paper, we mainly consider the speed of converging to the steady state and focus on the deviation $\tilde{C}\left(  t\right)  =C\left(  t\right)  -C\left(\infty\right)$, whose
evolution equation is given by
\begin{equation}
\frac{d}{dt}\tilde{C}\left(  t\right)  =X\tilde{C}\left(  t\right)
+\tilde{C}\left(  t\right)  X^{\dagger},
\end{equation}
where the damping matrix $X=i( H_{0}^{T}+i M ) $ and directly connects to the non-Hermitian Hamiltonian $H$ in real space as $X = iH^*$.
Here, $M$ is a diagonal matrix whose diagonal 
elements are $\{0,\gamma_1,0,\gamma_2,\cdots,0,\gamma_n\}$. The solution of $\tilde{C}$ is
\begin{equation}\label{eq29}
\tilde{C}\left(  t\right)  =e^{Xt}\tilde{C}\left(  0\right) e^{X^{\dagger}t}.
\end{equation}
By expressing $X$ in terms of right and left eigenvectors,
\begin{equation}
X=\sum_{n}\lambda_{n}\left\vert \psi_{n}\right\rangle \left\langle \phi
_{n}\right\vert,
\end{equation}
we can write Eq.~(\ref{eq29}) as
\begin{equation}
\tilde{C}\left(  t\right)  =\sum_{n,n^{\prime}}\exp\left[  \left(
\lambda_{n}+\lambda_{n^{\prime}}^{\ast}\right)  t\right]  \left\vert \psi
_{n}\right\rangle \left\langle \phi_{n}\right\vert \tilde{C}\left(
0\right)  \left\vert \phi_{n^{\prime}}\right\rangle \left\langle
\psi_{n^{\prime}}\right\vert.
\end{equation}
Due to the dissipative nature of the system,  $\operatorname{Re}(\lambda_{n})\leq0$ always holds. The  Liouvillian gap is defined as $\Delta=\min\left[  2\operatorname{Re}\left(  -\lambda_{n}\right)  \right]$, which is crucial to the long-time dynamics.  A finite gap implies exponential
convergence towards the steady state, while a vanishing
gap implies algebraic convergence \cite{PhysRevLett.123.170401,PhysRevLett.111.150403}.

Using a method similar to Sec.~\ref{Sec:2}, we can demonstrate that the damping matrix $X$ of the open chain under PBC also has universal Liouvillian gapless points. We can check our conclusion numerically with a specific model. Let the hopping $t_m = 0, m=2,3,\cdots$, we can take the distribution of dissipators to be randomly distributed, namely $L_x=\sqrt{\gamma_x} c_x^B$, where $\gamma_x$ is a random number within $0.4$ and $0.6$, it varies for each $x$. And we get the Liouvillian spectrum [Fig.~\ref{FIG8}(b)]. We can see that if the condition of Eq.~(\ref{eq:connection_condition}) can be satisfied, the Liouvillian spectrum can have Liouvillian gapless points, which correspond to the eigenmodes of the form $\left(1,0,e^{ik},0,e^{i2k},0,\cdots,e^{ikN},0\right)$. These modes can be viewed as dark modes, which lead to the algebraic convergence towards the steady state in the bulk [Fig.~\ref{FIG8}(c)]. Here, we already replaced the loss probability on B sites $P_x$ with the steady-state density on B sites $n^B_x$,  which can be written as \cite{PhysRevB.108.235422}
\begin{equation}
    n_x^B=[C(\infty)]_{xB,xB}=\gamma_x\int_{-\infty}^{+\infty}\frac{d\omega}{\pi}\left| \braket{xB\left| \frac{1}{i\omega-X}\right|x_0 A} \right|^2.\label{eq:pxgf2}
\end{equation}
Comparing Eq.~(\ref{eq:pxgf}) and Eq.~(\ref{eq:pxgf2}), we can find the equivalence between the loss probability $P_x$ and the steady-state density $n^B_x$. 
Results show that the Liouvillian gapless points are indeed only related to the properties of Hermitian Hamiltonian under PBC and irrelevant to the dissipation of the system.

\section{Summary}\label{Sec:6}
In summary, our investigation centers on IGC points and relative dynamic characteristics in dissipative systems. For the onsite dissipative non-Hermitian model, we deduce the equation governing the IGC points. This analysis reveals that IGC states are unaffected by the non-Hermitian part of the Hamiltonian and serve as dark modes characterized by the absence of a population on dissipative sites. Then, we examine quantum walks across diverse scenarios and parameters for a class of one-dimensional dissipative chains, highlighting the pivotal role of IGC points under PBC in shaping quantum walk dynamics. Finally, we show that the PBC damping matrix of the dissipative chain can exhibit Liouvillian gapless points, which correspond to an algebraic convergence towards the steady state in long-time dynamics. As the onsite dissipation is feasible to implement, our theory of IGC points can be confirmed in various non-Hermitian platforms, for example, the photon quantum walk\cite{PhysRevLett.126.230402}, dissipative cold atom systems\cite{PhysRevLett.124.070402}, and nuclear spin syetems\cite{PhysRevResearch.3.013256}.

\begin{acknowledgments}
The authors would like to thank Pengyu Wen, and Yunyao Qi for their helpful discussion. 
This work is supported by the National Natural Science Foundation of China under Grants No.~11974205, and No.~61727801, 
and the Key Research and Development Program of Guangdong province (2018B030325002).
\end{acknowledgments}

\nocite{*}

\bibliography{apssamp}

\end{document}